\documentclass[twocolumn,showpacs,preprintnumbers,amsmath,amssymb]{revtex4}
\usepackage{graphicx}% Include figure files
\usepackage{dcolumn}% Align table columns on decimal point
\usepackage{bm}% bold math

\begin{document}

\title{Temperature and Pressure Dependence of the Fe-specific Phonon Density of States in Ba(Fe$_{1-x}$Co$_x$)$_2$As$_2$.}

\author{O.~Delaire$^1$}
\author{M. S.~Lucas$^1$}
\author{A.M.~dos Santos$^1$}
\author{A.~Subedi$^{1,2}$}
\author{A.S.~Sefat$^1$}
\author{M.A.~McGuire$^1$}
\author{L.~Mauger$^3$}
\author{J.A.~Mu\~noz$^3$}
\author{C.A.~Tulk$^1$}
\author{Y.~Xiao${}^{4}$}
\author{M.~Somayazulu$^5$}
\author{J.Y.~Zhao$^6$}
\author{W.~Sturhahn$^6$}
\author{E.E.~Alp$^6$}
\author{D.J.~Singh$^1$}
\author{B.C.~Sales$^1$}
\author{D.~Mandrus$^1$}
\author{T.~Egami${}^{1,2}$}

\affiliation{ 1. Oak Ridge National Laboratory, 1, Bethel Valley road, Oak Ridge, TN 37831,  USA \\
2. Department of Physics and Astronomy, University of Tennessee, Knoxville, TN 37996, USA\\
3. California Institute of Technology,W. M. Keck Laboratory 138-78, Pasadena, CA 91125, USA \\
4. HPCAT, Geophysical Laboratory, Carnegie Institution of Washington, Argonne, IL 60439, USA\\
5. Geophysical Laboratory, Carnegie Institution for Science, Washington, D.C., 20015, USA \\
6. Advanced Photon Source, Argonne National Laboratory, Argonne, IL 60439, USA \\
}

\date{\today}

\begin{abstract}
The $^{57}$Fe-specific phonon density of states of Ba(Fe$_{1-x}$Co$_x$)$_2$As$_2$ single crystals ($x=0.0, 0.08$) was measured at cryogenic temperatures and at high pressures with nuclear-resonant inelastic x-ray scattering. Measurements were conducted for two different orientations of the single crystals, yielding the orientation-projected $^{57}$Fe-phonon density of states (DOS) for phonon polarizations in-plane and out-of-plane with respect to the basal plane of the crystal structure. In the tetragonal phase at 300$\,$K, a clear stiffening was observed upon doping with Co. Increasing pressure to $4\,$GPa caused a marked increase of phonon frequencies, with the doped material still stiffer than the parent compound.
Upon cooling, both the doped and undoped samples showed a stiffening, and the parent compound exhibited a discontinuity across the magnetic and structural phase transition. These findings are generally compatible with the changes in volume of the system upon doping, increasing pressure, or increasing temperature, but an extra softening of high-energy modes occurs with increasing temperature. First-principles computations of the phonon DOS were performed and showed an overall agreement with the experimental results, but underestimate the Gr\"uneisen parameter. This discrepancy is explained in terms of a magnetic Gr\"uneisen parameter, causing an extra phonon stiffening as magnetism is suppressed under pressure.

\end{abstract}

%\pacs{64.75.+g, 63.70.+h, 61.12.-q}% PACS, the Physics and Astronomy

\maketitle

\section{Introduction}

The recent discovery of superconductivity in iron pnictides at temperatures up to $55\,$K has generated great excitement, and many investigations have sought to  clarify the mechanism responsible for superconductivity in these compounds. Although this new class of compounds presents some similarities with the cuprate superconductors (\textit{e.g.} layered structure and nearness to magnetism), the parent compounds in the case of the iron pnictides are metals, rather than insulating oxides. Early first-principles calculations of the electron-phonon coupling have shown that a conventional electron-phonon coupling in the Bardeen-Cooper-Schrieffer theory is too weak to explain the magnitude of the observed superconducting temperature, $T_c$ \cite{Boeri}. Measurements of phonons in several compounds have been performed, but no significant changes in the phonon density of states (DOS) have been reported across $T_c$ \cite{Christianson, Phelan}. However, an isotope effect was observed in Ba$_{1-x}$K$_{x}$Fe$_2$As$_2$ and SmFeAsO$_{1-x}$F$_{x}$ by Liu {\it et al.}\cite{Liu-isotope}, while Shirage {\it et al.} observed an inverse isotope effect in Ba$_{1-x}$K$_{x}$Fe$_2$As$_2$ \cite{Shirage-inverse-isotope}.
Also, some anomalies have been in reported in the behavior of phonons \cite{Mittal-CaFe2As2}, and recent reports have emphasized the potential importance of coupling of phonons to the magnetic structure, or to spin fluctuations \cite{Yildirim, Zbiri, Reznik-BaFe2As2-IXS, Egami-spin-phonon, Hahn, Fernandes}.

Superconductivity in the iron pnictides can be achieved by doping the parent compounds either in the Fe planes themselves, as in the present case of Ba(Fe$_{1-x}$Co$_x$)$_2$As$_2$ \cite{Athena-PRL}, or in the separating layers, as in Ba$_{1-x}$K$_{x}$Fe$_2$As$_2$ or LaO$_{1-x}$F$_{x}$FeAs. Superconductivity can also be achieved in the undoped parent compounds by increasing pressure \cite{Torikachvili, Park, Alireza, Colombier}, although the degree of hydrostaticity may influence the appearance of superconductivity \cite{Matsubayashi}. Several studies have shown that the magnetism in BaFe$_2$As$_2$ is suppressed with increasing pressure \cite{Colombier, Matsubayashi, Fukazawa}, and that the parent compound becomes superconducting, with a maximum in $T_c$ at $P \simeq 4\,$GPa under quasi-hydrostatic conditions (as obtained in diamond anvil cells) \cite{Alireza, Colombier}. 

Changes in composition, temperature, and pressure can all lead to shifts in the phonon energies of a crystal, but these are often interrelated. For example, temperature will affect volume through thermal expansion. Doping also commonly leads to changes in lattice parameters, besides changing {\textit e.g.} carrier concentration. As a result, the effects of temperature alone (at constant volume), or doping alone, is not straightforward to establish, unless one records the dependency on the multiple coupled parameters. Using Nuclear Resonant Inelastic X-ray Scattering (NRIXS), we have performed a systematic study of phonons in Ba(Fe$_{1-x}$Co$_x$)$_2$As$_2$, as function of temperature, pressure, and doping, in order to compare the effect of different thermodynamic parameters on the phonon DOS.

Previous investigations have not reported a systematic effect of doping on the phonons in Ba(Fe$_{1-x}$Co$_x$)$_2$As$_2$, although effects of doping on phonons have been reported in other iron-pnictides \cite{LeTacon-NdFeAsO, LeTacon-SmFeAsO,Lee-phonons-BaKFe2As2}. We show in the present study that doping in BaFe$_2$As$_2$ also has an  affect on the phonons. The effect of an increase in pressure is directly observed in the phonons, and the effect of compression has some similarity with the effect seen upon doping. The temperature dependence of the phonons can also be interpreted in terms of the change in the lattice volume due to thermal expansion, and we could compare the predicted effect of thermal expansion within the quasi-harmonic approximation, to the effect observed upon compression.

\section{Sample Synthesis}

Single-crystalline samples of BaFe$_2$As$_2$ and optimally-doped (for maximum $T_c$) Ba(Fe$_{1-x}$Co$_x$)$_2$As$_2$ ($x=0.08$) were synthesized via a flux-growth technique \cite{Athena-PRL}, using an Fe-As flux enriched in the $^{57}$Fe isotope. The isotopic enrichment was about 50\%. The crystals formed small shiny, smooth, plates about 2$\,$mm wide and 0.1$\,$mm thick.

\begin{figure}[tbp]
\center
\includegraphics[width = 7.0 cm]{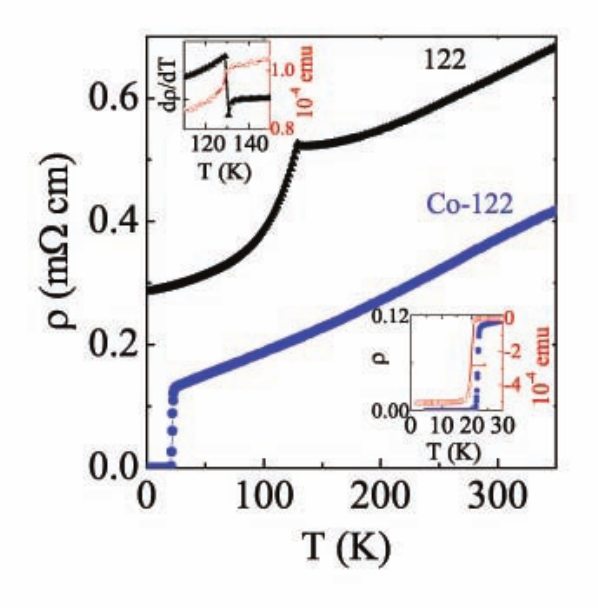} 
\caption{(Color online) Resistivity curves for BaFe$_2$As$_2$ (122) and BaFe$_{1.84}$Co$_{0.16}$As$_2$ (Co-122) synthesized with 50\% $^{57}$Fe-enriched iron. The top inset shows the derivative of the resistivity, and the magnetization, for the parent compound BaFe$_2$As$_2$. The bottom inset shows the resistivity and magnetization of BaFe$_{1.84}$Co$_{0.16}$As$_2$ at low temperature. 
} \label{resistivity}
\end{figure}

Resistivity and magnetic susceptibility measurements were performed on the $^{57}$Fe-enriched samples. Results are shown in Fig.~\ref{resistivity}. The spin-density wave transition temperature of the parent compound was $T_{\rm SDW}=130\,$K, and the superconducting transition temperature in the doped compound was $T_c = 22\,$K.  There was no detectable shift in the superconducting transition temperature with partial $^{57}$Fe enrichment in our samples. Liu {\textit et al.} have reported a positive isotope effect for $^{54}$Fe enrichment in Ba$_{1-x}$K$_{x}$Fe$_2$As$_2$ \cite{Liu-isotope}. However, Shirage {\textit et al.} reported an inverse isotope effect in the same compound with $^{54}$Fe and $^{57}$Fe isotopes \cite{Shirage-inverse-isotope}, so a definite isotope effect remains to be established.

X-ray diffraction patterns ($\lambda=1.54056 \rm \AA$)  were acquired at 300$\,$K on powders obtained by crushing the platelets. The x-ray patterns were refined with a Rietveld procedure, as implemented in the software FULLPROF \cite{fullprof}. Results are listed in Table~\ref{XRD}. 
The substitution of Fe by Co leads to a slight contraction of $c$ parameter ($-0.27$\%), while the $a$ and $b$ parameters are almost unchanged (difference smaller than 0.03\%). The contraction of the $c$-axis upon doping and the lack of change for $a$ and $b$ axes are consistent with the lattice parameters previously reported for the case of samples synthesized from natural iron \cite{Athena-PRL}.

% XRD refinements
\begin{table}[tbp]
\caption{\label{XRD}Rietveld refinements of crystal structure of $^{57}$Fe-enriched samples ($T=300\,$K, $\lambda=1.54056 \rm \AA$). }
\begin{ruledtabular}
\begin{tabular}{lccc}
{sample}  &  BaFe$_2$As$_2$    &    BaFe$_{1.84}$Co$_{0.16}$As$_2$    \\
 \hline
$a=b \,(\rm \AA)$  & $3.9613(1)$       & $3.9604(1)$    \\
$c \,(\rm \AA)$  & $13.0148(3)$ & $12.9793(3)$ \\
$z_{\rm As}$ \footnotemark[1] &  $0.35423(7)$ & $0.35423(8)$
\footnotetext[1]{Fractional coordinate.} 
\end{tabular}
\end{ruledtabular}
\end{table}

\section{Nuclear-Resonant Inelastic X-ray Scattering Measurements}

Nuclear resonant inelastic x-ray scattering (NRIXS) measurements~\cite{Alp-NRIXS, Seto, Sturhahn-INRS} were performed at cryogenic temperatures at beamline 3-IDD and high pressures at beamline 16-IDD (HP-CAT) at the Advanced Photon Source at the Argonne National Laboratory. 

\begin{figure}[tbp]
\center
\includegraphics[width = 8.0 cm]{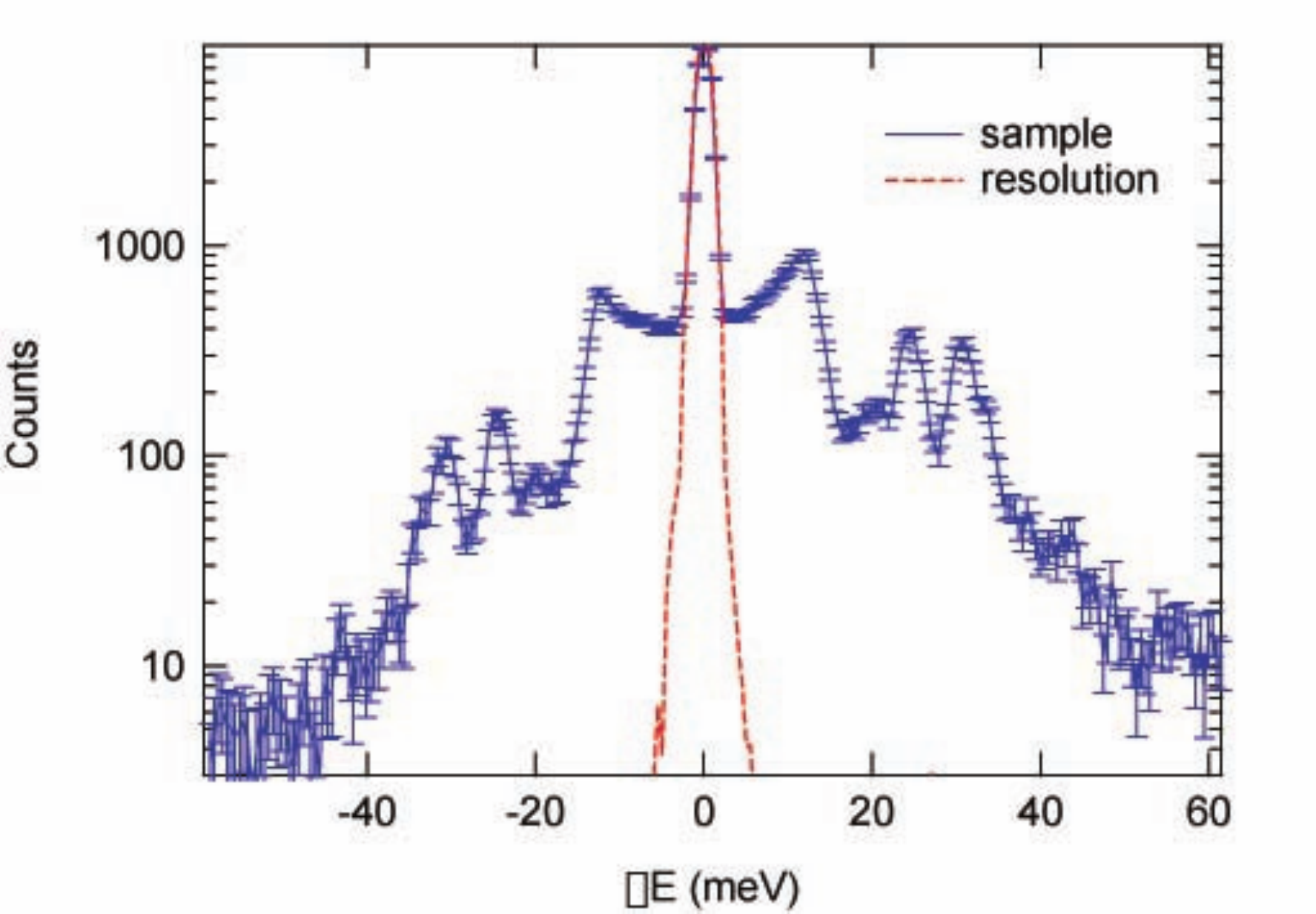} 
\caption{(Color online) NRIXS data measured on BaFe$_2$As$_2$ sample at $T=300\,$K and $P=0\,$GPa (solid line, ``sample''), and measured resolution function of beamline 16-IDD (dashed line, ``resolution''). $\Delta E$ is the shift from the $^{57}$Fe nuclear resonance energy.} \label{resolution}
\end{figure}

In all measurements, the incident photon energy was tuned to 14.4124$\,$keV, the nuclear resonance energy of $^{57}$Fe. The NRIXS signal was measured with multiple avalanche photodiode detectors positioned 90$^{\circ}$ from the direction of the beam, with the exception of the IP high pressure measurement where detectors were positioned 60$^{\circ}$ from the direction of the beam (due to pressure cell geometry constraints). Data were collected in scans of incident photon energy, with  $\Delta E = -60$ to $+60\,$meV from the resonant energy, in steps of $0.5\,$meV. The experimental energy resolution function was measured with a single avalanche photodiode placed in the forward beam direction, recording the intensity as a function of the shift of the incident energy away from the $^{57}$Fe resonance energy (the data for the instrument resolution were summed over all runs performed in the same conditions).
Representative spectra for the NRIXS data and the instrumental resolution are shown in Fig.~\ref{resolution}. The monochromator energy resolution (full width at half maximum) was $2.2\,$meV in the measurements as function of pressure, and $2.5\,$meV in the measurements as function of temperature.

All of the NRIXS data reduction were performed using
the standard software PHOENIX  \cite{Sturhahn}.
The raw NRIXS spectra, given as intensity versus the
angle of the monochromator crystals, were converted to intensity versus
energy transfer. The first few bins on the low energy
side (-60meV to -55meV) were used to determine
an energy independent background, and was removed for
all energy transfers. The elastic peak was removed using
the measured resolution function.
The contribution from multiphonon scattering processes
was subtracted using a self-consistent procedure based on
a Fourier-log method \cite{Johnson}, and the Fe-partial phonon
density of states was obtained by correcting for the thermal
occupation factor.

\subsection{Pressure-Dependent Measurements}

Measurements as function of pressure were performed in panoramic diamond-anvil cells (DAC), with diamonds of 500$\, \mu$m-diameter culets. The samples were 100$\,\mu$m-wide plate-like crystals, and were contained inside the DAC's by Be gaskets drilled with 130$\, \mu$m holes (platelets remained parallel to the gasket plane). The pressure medium was silicone oil, which provides a nearly hydrostatic pressure medium up to $10\,$GPa \cite{Klotz-pressure-media}. The pressure inside the DAC's was determined through the fluorescence line of ruby crystals loaded with the samples in the pressure medium \cite{Piermarini_1975}. Measurements were performed with the cells at ambient pressure ($0\,$GPa) and with the cells pressurized to $4.0\pm0.2\,$GPa. 

\begin{figure}[htbp]
\center
\includegraphics[width = 7.5 cm]{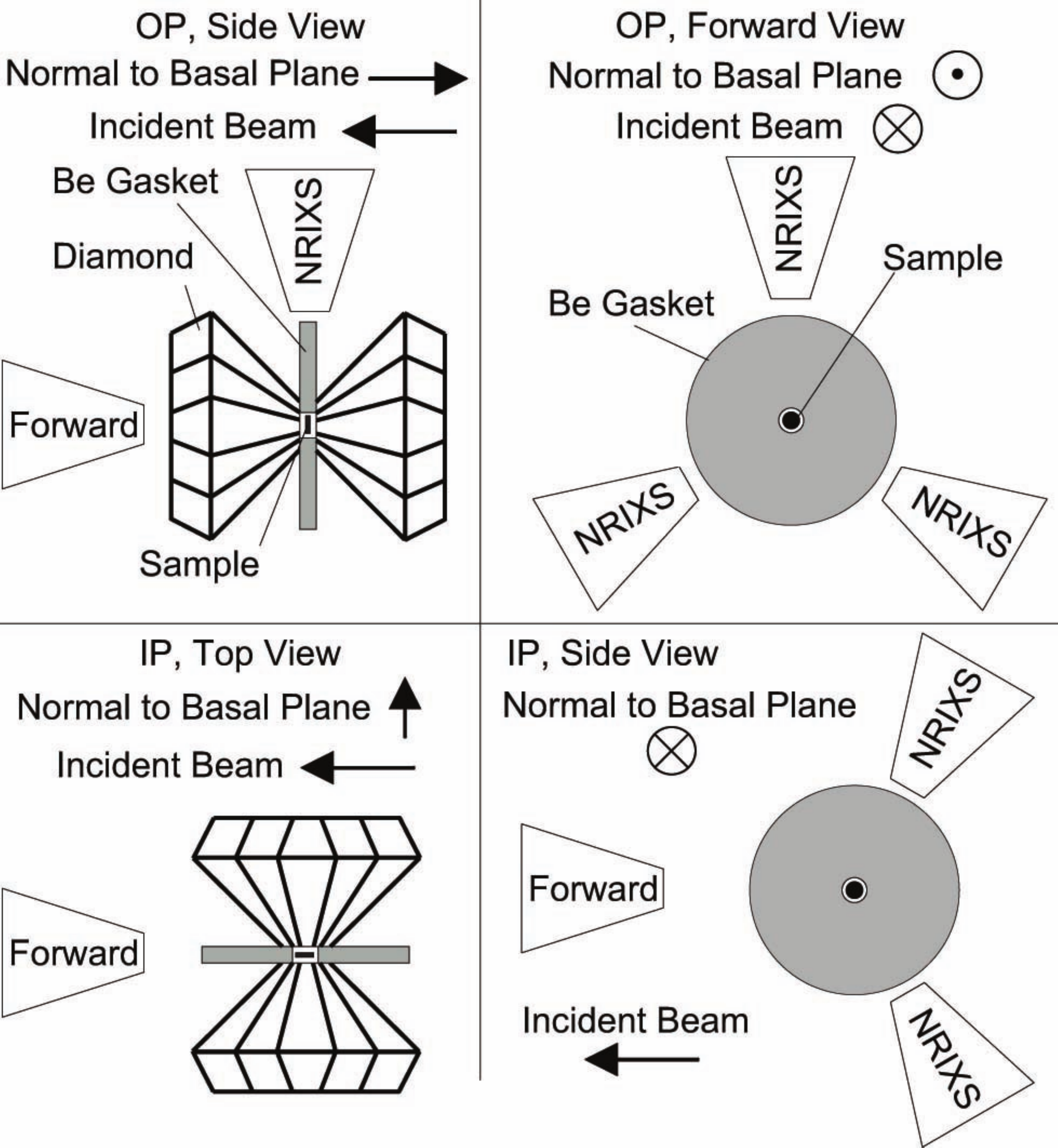}
\caption{Schematic of geometry of high pressure NRIXS measurements.  See text for description.} \label{orientation}
\end{figure}

Out-of-plane (OP) measurements were performed with the incident beam passing through the diamonds and the scattered NRIXS signal measured through the Be gasket.  In-plane (IP) measurements were performed with the incident beam passing through the Be gasket and the scattered NRIXS signal also measured through the Be gasket.  The geometry of the high-pressure NRIXS measurements is illustrated in Fig.~\ref{orientation}.

\subsection{Temperature-Dependent Measurements}

Measurements as function of temperature (10$\,$K to 300$\,$K, at ambient pressure) were performed in an evacuated He-flow cryostat, with the sample mounted on a Cu cold finger and exposed to the x-ray beam through a beryllium dome. The temperature was monitored with two thermocouples positioned respectively on the cryostat head, and directly next to the sample on a Cu bracket. The readings from the two thermocouples were in good agreement throughout the measurements.

\section{Experimental Results}

\subsection{Effect of Doping}

\begin{figure}[tbp]
\center
\includegraphics[width = 8.0 cm]{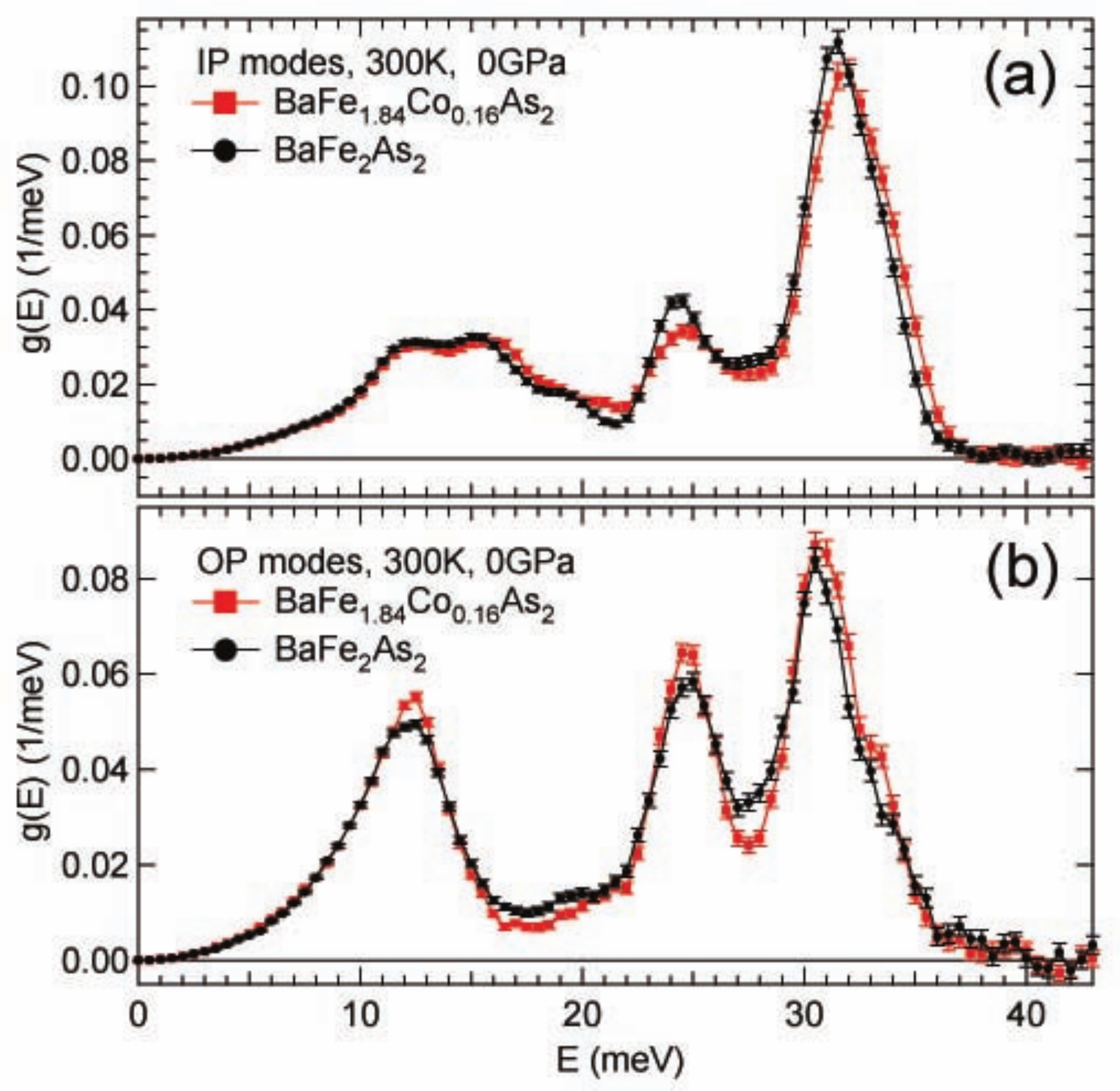} 
\caption{(Color online) (a) Fe-partial phonon DOS for modes with in-plane (IP) polarizations in Ba(Fe$_{1-x}$Co$_x$)$_2$As$_2$ for $x=0.0$ and $x=0.08$, at ambient conditions ($T=300\,$K, $P=0\,$GPa). (b) Same for out-of-plane (OP) polarizations. } \label{doping_300K_0GPa}
\end{figure}

The Fe-partial phonon DOS ($300\,$K, $0\,$GPa) for IP and OP modes in BaFe$_2$As$_2$ and BaFe$_{1.84}$Co$_{0.16}$As$_2$ are shown in Fig.~\ref{doping_300K_0GPa}. The Fe-partial phonon DOS is composed of three peaks around 13, 25, and 32$\,$meV. The low and medium-energy peaks are more prominent in the OP-polarized DOS, indicating that a larger proportion of displacements along the $c$ axis in these energy ranges.  This is physically intuitive, as the crystal structure is expected to have lower vibration frequencies for displacements of Fe atoms along $c$.  In the parent compound ($x=0$), we find for the first-moment of the phonon energies: $\langle E^{x=0}_{\rm IP} \rangle = 24.57 \pm 0.04\,$meV for IP vibrations, while $\langle E^{x=0}_{\rm OP} \rangle = 22.93 \pm 0.06\,$meV for OP vibrations. %The high-energy peak in the IP DOS corresponds to the phonon bands around the high-energy $E_g$ Raman mode (in plane Fe and As motions) \cite{Chauviere}. 

As can be seen in this Fig.~\ref{doping_300K_0GPa}, the phonons are stiffer in the doped material than in the parent compound, for both the IP and OP-polarized vibrations. This is particularly well seen on the peak at about 32$\,$meV and at the phonon cutoff. The mass difference of Co and Fe atoms cannot account for this shift, since Co is more heavy than Fe. The phonon stiffening is about $0.5\,$meV at the cutoff ( $\sim1.5$\% change), which is significant, considering that doping only causes a modest contraction in the $c$ lattice parameter of order 0.25\%, and almost no change in the $a$ and $b$ parameters. The stiffening of phonons upon doping was also observed in the measurements at $4\,$GPa, where a similar magnitude in the shift upon doping persisted. 
Our analysis of the cutoff energy gives $\Delta E^{\rm IP}_{\rm cut}=0.5\pm0.1\,$meV for IP modes, and $\Delta E^{\rm OP}_{\rm cut}=0.4\pm0.1\,$meV for OP modes, at $300\,$K and $0\,$GPa. 

\subsection{Effect of Pressure}

\begin{figure}[tbp]
\center
\includegraphics[width = 8.0 cm]{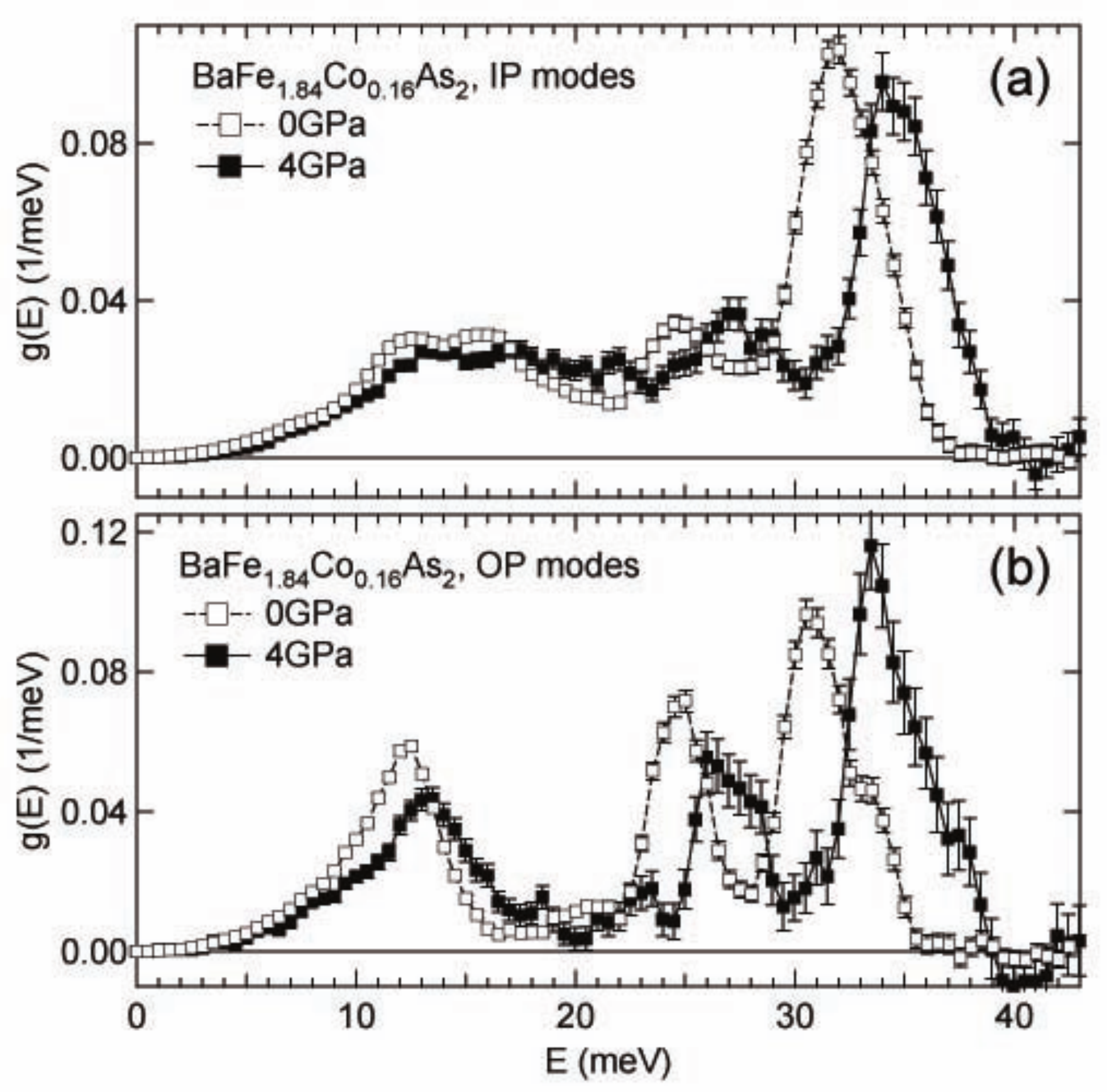} 
\caption{(a) Partial Fe phonon DOS for in-plane (IP) polarized modes in BaFe$_{1.84}$Co$_{0.16}$As$_2$, at $0\,$GPa and $4.0\pm0.2\,$GPa ($300\,$K). (b) Same for out-of-plane polarizations (OP).} \label{DOS_BFCA_vs_P}
\end{figure}

\begin{figure}[tbp]
\center
\includegraphics[width = 8.0 cm]{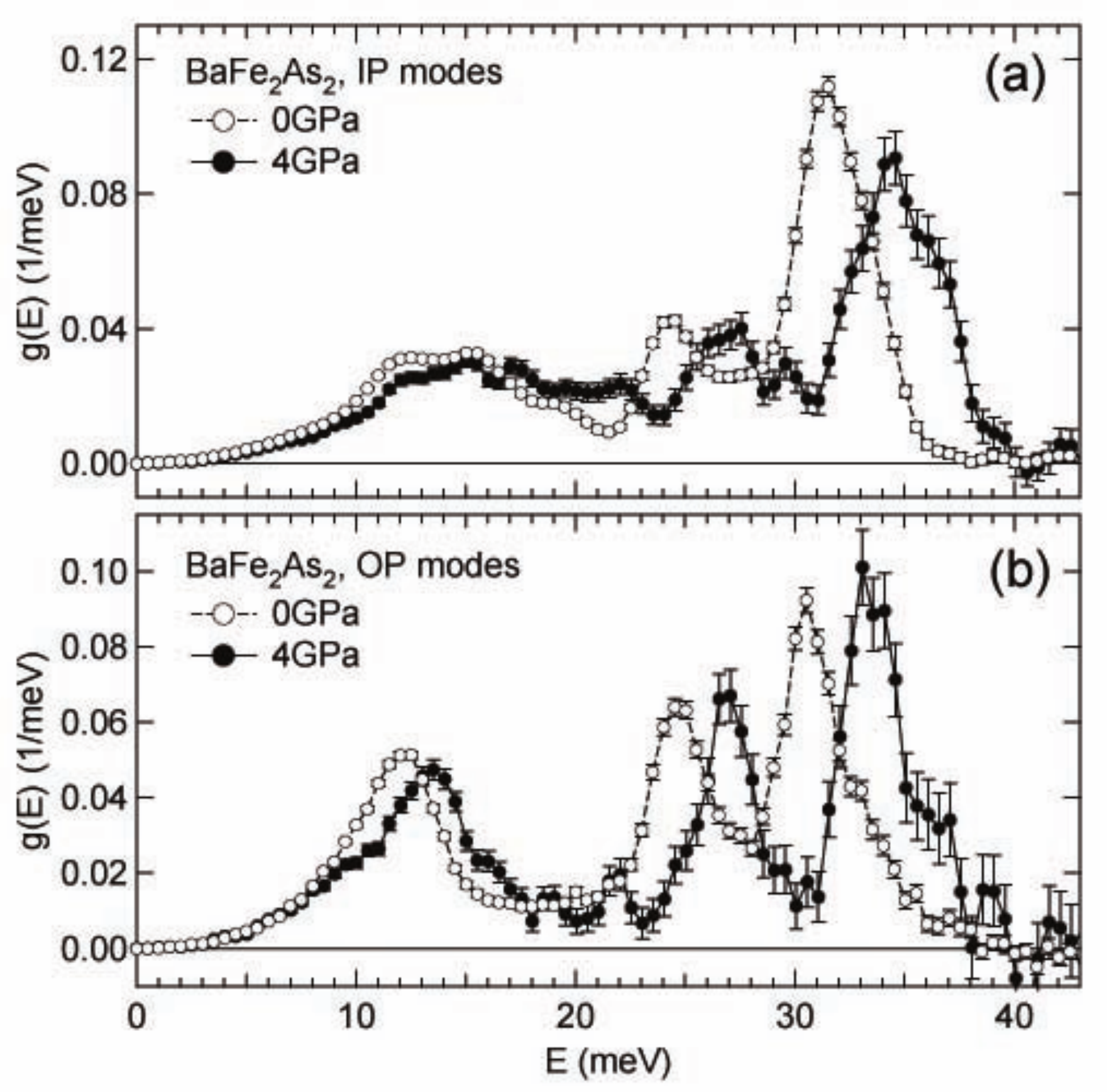} 
\caption{(a) Partial Fe phonon DOS for in-plane (IP) polarized modes in BaFe$_2$As$_2$, at $0\,$GPa and $4.0\pm0.2\,$GPa ($300\,$K). (b) Same for out-of-plane polarizations (OP).}
 \label{DOS_BFA_vs_P}
\end{figure}

Measurements under pressure were performed with single crystals loaded inside the diamond anvil cells. The single crystals were oriented with the $c$-axis parallel to the axis of the DAC. This orientation was verified using in-situ diffraction (both at $0\,$GPA and $4\,$GPa), and also confirmed by comparing the phonon spectra at $0\,$GPa with samples in the DACs and the measurements without the pressure cells. Area diffraction patterns indicated some turbostratic disorder in the $(a,b)$ plane of the samples (rotational stacking disorder along $c$), but this type of disorder does not hinder the measurement of separate in-plane and out-of-plane polarizations with our setup.  The results for the Fe-partial phonon DOS of BaFe$_{1.84}$Co$_{0.16}$As$_2$ and BaFe$_2$As$_2$ are shown in Figs.~\ref{DOS_BFCA_vs_P} and \ref{DOS_BFA_vs_P}, respectively.

The effect of pressure on the phonon DOS is significant. In both the parent and the doped samples, a large stiffening is observed. The pressure dependence of the in-plane (IP) modes is similar in doped and undoped materials, but the out-of-plane modes (OP) appear more sensitive to pressure in the doped material. This may be due to the doping-induced $c$-axis reduction in the superconductor. From the slopes, we find the pressure coefficients: $d \langle E_{\rm OP}  \rangle / dP =  0.68\,$ and $0.45\,$meV/GPa for OP modes in doped and parent compounds, respectively, and $d \langle E_{\rm IP}  \rangle / dP =  0.40\,$ and $0.43\,$meV/GPa for IP modes in doped and parent compounds, respectively.

\begin{figure}[tbp]
\center
\includegraphics[width = 6.5 cm]{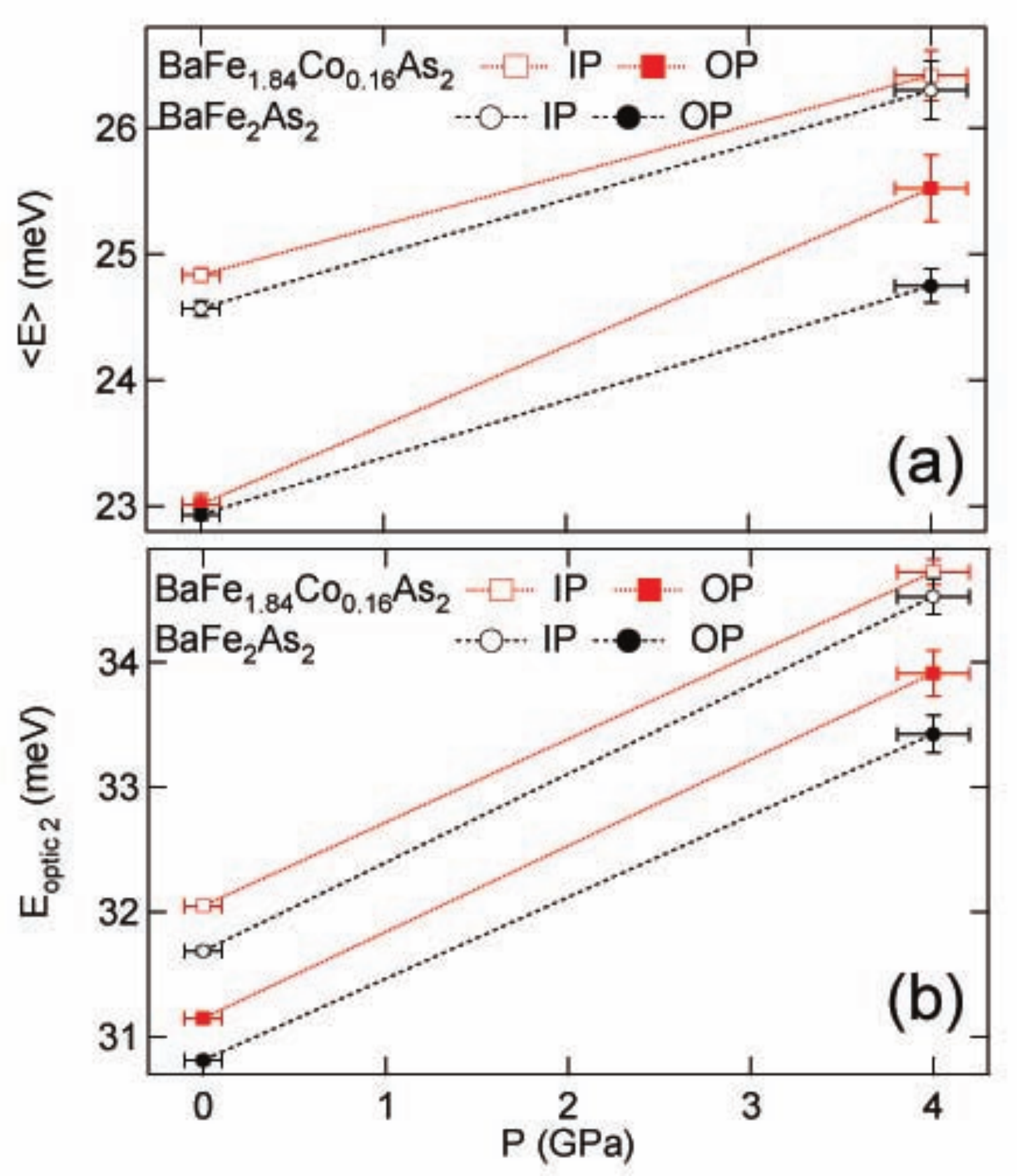} 
\caption{(Color online) (a) Average energy as function of pressure in BaFe$_2$As$_2$ and BaFe$_{1.84}$Co$_{0.16}$As$_2$ (300$\,$K). (b) Energy of top optical peak as function of pressure in BaFe$_2$As$_2$ and BaFe$_{1.84}$Co$_{0.16}$As$_2$ (300$\,$K).  } \label{Eavg_vs_P}
\end{figure}

Our in-situ diffraction data on BaFe$_{1.84}$Co$_{0.16}$As$_2$ and BaFe$_2$As$_2$ at $4\,$GPa show lattice compressions compared to $0\,$GPa that are in good agreement with those reported by Kimber \textit{et al.} \cite{Argyriou-pressure} (although we could not resolve in-situ the difference in $c$-parameter between the two compounds). We find the following lattice parameters for BaFe$_{1.84}$Co$_{0.16}$As$_2$ at $4\,$GPa and $300\,$K:
$a = b = 3.90\pm0.01\,\rm \AA$ ; $c = 12.71\pm0.01\,\rm \AA$. 
Within the error bars of our \textit{in-situ} measurements, the same values are obtained for BaFe$_2$As$_2$ at 4GPa and 300K. Both the superconductor and the parent compound  appear to compress by the same relative amount between $0\,$GPa and $4\,$GPa: $-2.5\pm0.1$\% for $c$, and about $-1.5\pm0.2$\% for $a,b$  (all at 300K). Kimber \textit{et al.} have reported that the compression of $a$ and $c$ axes is very linear with pressure from $1\,$GPa to $6\,$GPa, and that compressibilities along both axes are unchanged between 17 and $150\,$K. Our results indicate that this also holds at $300\,$K. 

From the experimental change in volume,  and the measured phonon frequencies, we obtained the mode-dependent Gr\"uneisen parameters at $T=300\,$K, $\gamma = -d \ln E / d \ln V$, for the average phonon energy as well as the high-energy optical modes. Results are listed in Table~\ref{Gruneisen-exp}.

% Gruneisen parameters
\begin{table}[tbp]
\caption{\label{Gruneisen-exp}Experimental Gr\"uneisen parameters of  $^{57}$Fe-enriched Ba(Fe$_{1-x}$Co$_x$)$_2$As$_2$, obtained from NRIXS  measurements as function of pressure (T$=300\,$K) or temperature (P$=0\,$GPa). }
\begin{ruledtabular}
\begin{tabular}{lccc}
{}  & {$\gamma_{\rm T=300\,K}$} & {$\gamma_{\rm T=300\,K}$}  & {$\gamma_{\rm P=0\,GPa}$} \\
{polarization}  &  IP    &  OP  &  OP    \\
 \hline
$\langle E \rangle$ & {} & {} & {} \\
BaFe$_2$As$_2$  &   $1.4\pm0.2$  &   $1.6\pm0.2$   &  $1.5\pm0.3$          \\
BaFe$_{1.8}$Co$_{0.2}$As$_2$   &   $1.3\pm0.2$  &  $2.1\pm0.2$  & $1.9\pm0.6$   \\
\hline
$E_{\rm optic,2}$  ($32\,$meV)  & {} & {}  & {}  \\
BaFe$_2$As$_2$  &   $1.8\pm0.1$  &   $1.7\pm0.1$  &  $2.2\pm0.2$   \\
BaFe$_{1.8}$Co$_{0.2}$As$_2$   &  $1.7\pm0.1$  &   $1.8\pm0.1$  &  $2.9\pm0.3$ 

\end{tabular}
\end{ruledtabular}
\end{table}

\subsection{Effect of Temperature}

\begin{figure}[tbp]
\center
\includegraphics[width = 8.0 cm]{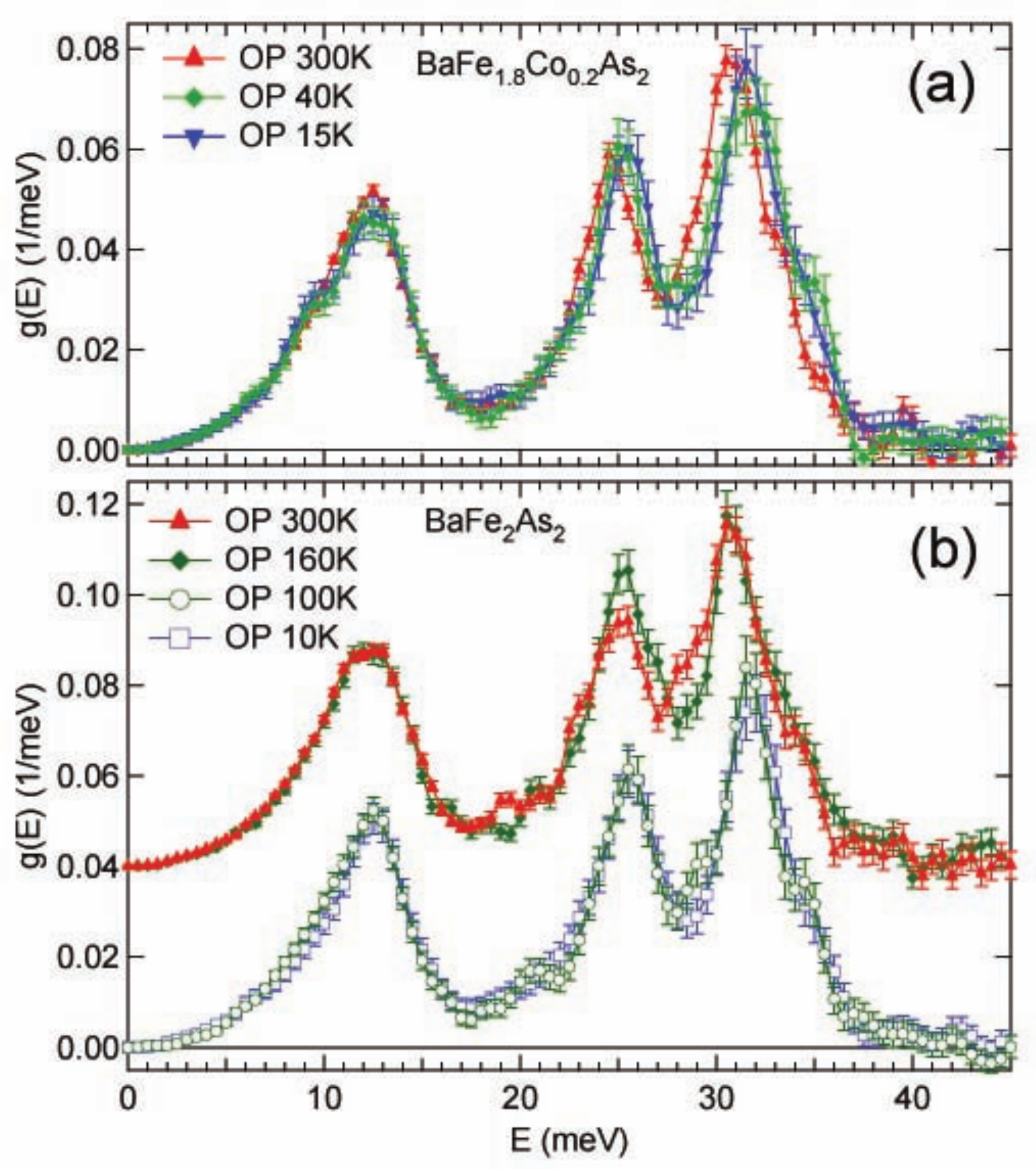} 
\caption{(Color online) (a) OP Fe-pDOS of BaFe$_{1.84}$Co$_{0.16}$As$_2$ as function of temperature, as measured with NRIXS. (b) Same for BaFe$_2$As$_2$. The curves for measurements at 160 and $300\,$K in the lower panel were vertically offset for clarity.} \label{DOS_BFA_BFCA_vs_T}
\end{figure}

The OP Fe-partial phonon DOS of BaFe$_{1.84}$Co$_{0.16}$As$_2$ was measured at 15, 40, 160, and 300$\,$K, and that of BaFe$_2$As$_2$ was measured at 10, 100, 160, and 300$\,$K. We do not observe any large change in the OP Fe-pDOS of BaFe$_{1.84}$Co$_{0.16}$As$_2$ across $T_{c} =22\,$K. The main effect observed is an overall softening of the phonon DOS with increasing $T$. This softening corresponds to a downward shift of about 1$\,$meV in the position of the $32\,$meV peak between $10\,$K and $300\,$K, compatible with the effect observed on the same peak in LaO$_{0.9}$F$_{0.1}$FeAs \cite{Christianson}. As we show below, this behavior is compatible with the quasiharmonic (QH) model, according to which phonon energies depend on temperature through the thermal expansion: $\Delta E(T) / E_0 = -\gamma \,  \Delta V(T) / V_0$, with $\gamma$ the Gr\"uneisen parameter. 

The parent compound also shows a systematic decrease in energy of all the modes upon heating from 10$\,$K to 300$\,$K, by an amount comparable to that in the doped material. The phonon DOS in the tetragonal and orthorombic phases are similar overall, as expected from the small magnitude of the structural distortion across the transition \cite{Rotter, Huang}. However, we do observe a slight change in shape of the DOS across the tetragonal--orthorombic  phase transition at $130\,$K. The optical phonon peak at $\sim32\,$meV appears to be the most sensitive, and shifts down abruptly across the transition (heating). Additionally, the phonon peak at $12\,$meV appears more broad above the phase transition. 

\begin{figure}[tbp]
\center
\includegraphics[width = 7.0 cm]{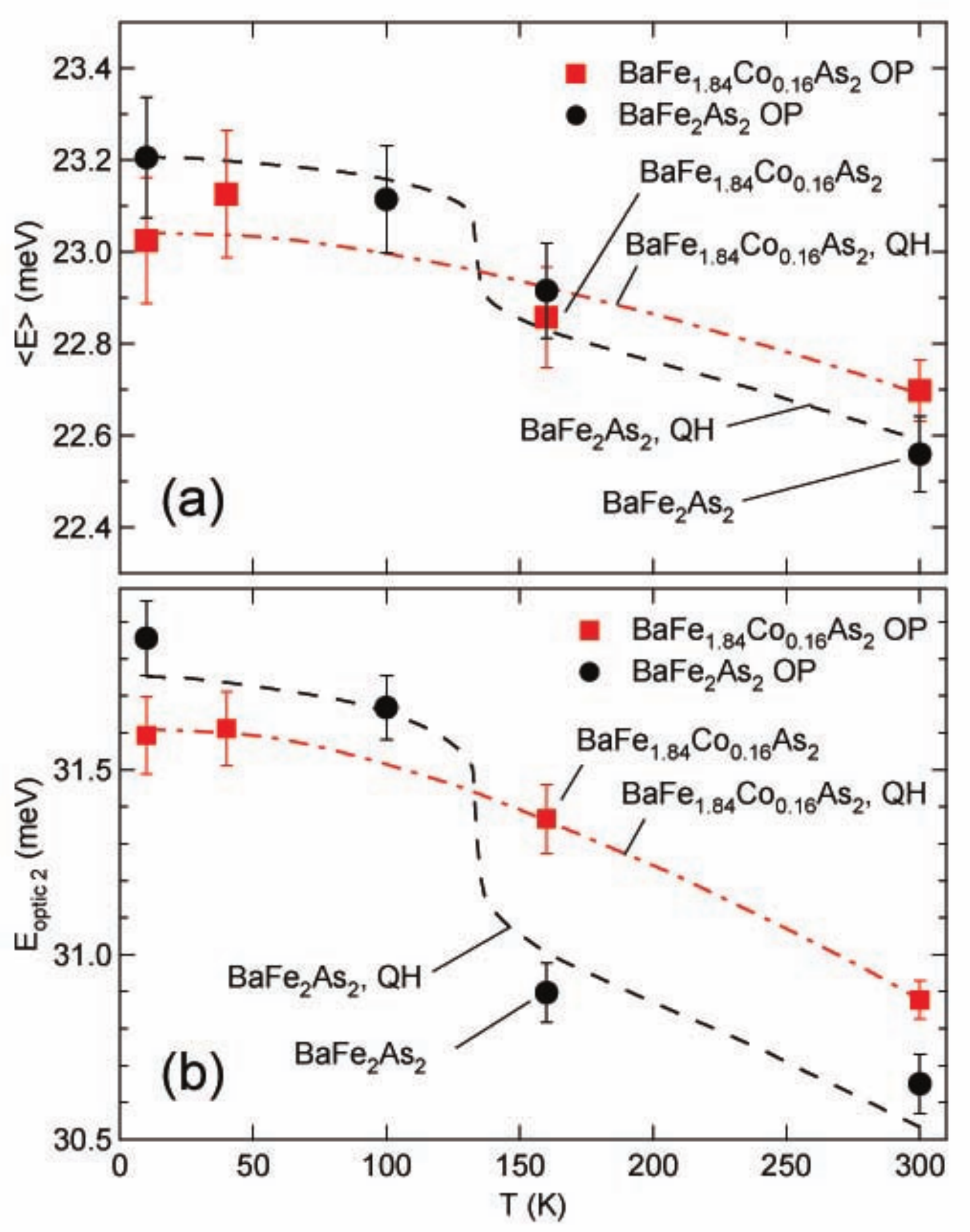} 
\caption{(Color online) (a) Temperature dependence of average energy for Fe vibrations with out-of-plane polarization (OP), in BaFe$_2$As$_2$ and BaFe$_{1.84}$Co$_{0.16}$As$_2$. (b)  Same for OP high-energy optic peak. Markers with error bars are experimental values. Dashed curves give the calculated $T$-dependence for a quasiharmonic  (QH) model, $E_{\rm QH}(T) = E_{0} \big( V(T) / V_{0} \big)^{-\gamma}$, using thermal expansion data reported in \cite{Budko-expansion, Daluz-expansion}, see text.} \label{Eavg_vs_T}
\end{figure}

The temperature dependence of the average energy, and of the high-energy optical phonon (around $32\,$meV) for OP modes is plotted in Fig.~\ref{Eavg_vs_T}. The average energy $\langle E \rangle$ was obtained from the first moment of the measured DOS, and the high-energy optical peak position was estimated by fitting the top half of the peak to a Gaussian. Error bars are from experimental counting statistics. As can be seen on this figure, the phonon energy increases with decreasing temperature in both BaFe$_2$As$_2$ and BaFe$_{1.84}$Co$_{0.16}$As$_2$, as expected from the decrease in overall volume of the lattice at lower temperatures. The behavior in BaFe$_{1.84}$Co$_{0.16}$As$_2$ is smooth, and follows the behavior expected from QH theory.  Our data show the possibility of a small softening upon cooling across $T_{c}$ (see Fig.~\ref{Eavg_vs_T}), compatible with the upturn in volume expansion reported in \cite{Budko-expansion, Daluz-expansion}.  In the case of BaFe$_2$As$_2$, there is a discontinuity in the frequency with a jump at the temperature of the tetragonal-orthorombic (and magnetic) transition. This jump is $0.8\pm0.2\,$meV for the high-energy optic OP mode, with the orthorombic phase stiffer than the tetragonal phase.  Our observation can be related to the smaller volume of the orthorombic phase: Bud\'{}Ko \textit{et al.} have reported a volume collapse across $T_{\rm SDW}$, mostly associated to a decrease in the basal plane dimensions \cite{Budko-expansion}. When both the doped and undoped materials are in the tetragonal phase, the high-energy optic mode in the doped material is stiffer than in the undoped compound by about $0.5\,$meV. Below $T_{\rm SDW}$, the orthorombic BaFe$_2$As$_2$ is stiffer than tetragonal BaFe$_{1.84}$Co$_{0.16}$As$_2$.

We note that Choi et al. have reported Raman measurements in CaFe$_2$As$_2$, which show a $0.5\,$meV step-like stiffening of the $B_{1g}$ Raman mode (at $26\,$meV) when cooling across the structural transition in that compound \cite{Choi-Raman}. Our data around the energy of the $B_{1g}$ Raman mode ($26\,$meV) in BaFe$_2$As$_2$ are less clear than for the $31\,$meV mode, but show an energy difference of $0.3\pm0.2\,$meV between 160$\,$K and 100$\,$K. The measurements of Chauvi\`ere {\it et al.} for the $B_{1g}$ Raman mode in BaFe$_2$As$_2$ do not show a clear discontinuity either, but seem to show a kink around $T_{\rm SDW}$  \cite{Chauviere-Raman}. We do not observe any trace of the splitting of the low-energy $E_g$ Raman mode reported by these authors  \cite{Chauviere-Raman}, but this mode corresponds to vibrations in the $(a,b)$ plane (IP-polarized) \cite{Litvinchuk-Raman}. 
%Akrap {\textit et al.} report an anomalous drop of $\simeq 0.25\,$meV in the frequency of the infrared in-plane $E_u$ mode at $31\,$meV, reminiscent of the low-energy branch of the split low-energy $E_g$ Raman mode observed by Chauvi\`ere et al. The drop in the $E_u$ mode is small, however, and could potentially be due to the splitting of $a$ and $b$ lattice constants in the orthorombic phase.

The temperature dependence of the phonon frequencies can be computed from the thermal expansion of the crystal using the quasi-harmonic approximation. Using the temperature dependence of the cell volume reported by Bud\'{}Ko \textit{et al.} for the doped compound \cite{Budko-expansion}, and by Daluz \textit{et al.} for the parent compound \cite{Daluz-expansion}, we determined the Gr\"uneisen parameters giving the best fits to our measurements of the 32$\,$meV OP optical mode (corresponding to dashed curves in Fig.~\ref{Eavg_vs_T}).  Assuming a $T$-independent Gr\"uneisen parameter $\gamma = - d \ln E / d \ln V = {\rm const.}$, we optimized  $\gamma$ and the energy at low temperature, $E_{0}(T=10\, {\rm K})$, so that $E_{\rm QH}(T) = E_{0} \big( V(T) / V_{0} \big)^{-\gamma}$ would minimize the $\chi^2$ of  $E_{\rm QH}(T) - E_{\rm NRIXS}(T)$ for all measured temperatures.
Results are listed in Table~\ref{Gruneisen-exp}. We find $\gamma|_{P=0} = 2.2 \pm 0.2$ for BaFe$_2$As$_2$ and $\gamma|_{P=0} = 2.9 \pm 0.3$ for BaFe$_{1.84}$Co$_{0.16}$As$_2$ for the high-energy optical peak with OP polarization.  It is to be noted that these ``thermal'' Gr\"uneisen parameters are significantly larger than the pure volume Gr\"uneisen parameters obtained from measurements under pressure ($\gamma|_{T=300} \simeq 1.7$ for the high-energy OP peak), or from the values obtained from DFT calculations, as discussed below.

\section{First-Principles Simulations}

The first principle calculations of in-plane (IP) and out-of-plane
(OP) Fe-partial phonon DOS for BaFe$_2$As$_2$ were done within the framework
of density functional theory (DFT), with the generalized gradient
approximation of Perdew et al. \cite{Perdew-GGA} The phonon calculations were done
using linear response as implemented in Quantum Espresso \cite{Giannozzi}. We used experimental lattice parameters
when available but relaxed the As height via energy minimization. The basis
set cutoff for the wave functions was $40\,$Ry, while a $400\,$Ry cutoff was
used for the charge density. A $4 \times 4 \times 4$ special $k$-point grid was used for
Brillouin zone sampling. The dynamical matrices were calculated on a
$4 \times 4 \times 4$ grid and the dynamics was then interpolated on a much finer
$48 \times 48 \times 48$ grid to obtain the phonon DOS.

The lattice parameters used in the calculation were: $a=b=3.9625\, \rm \AA$, $c=13.0168\,\rm \AA$, corresponding to the experimental structure at ambient conditions ($300\,$K, $0\,$GPa) \cite{Rotter}. Relaxation of the As  fractional coordinate along $c$ gave $z_{\rm As}=0.34515$. For the comparison with measurements at $4\,$GPa, computations were performed on a compressed unit cell with $a = b= 3.913\,\rm \AA$, $c = 12.698\,\rm \AA$, corresponding the high-pressure diffraction measurements reported in \cite{Argyriou-pressure}, and also in good agreement with our experimental estimates. In this case, the relaxed $z_{\rm As}$ was $0.34786$. In order to compare the phonon DOS of the parent compound BaFe$_2$As$_2$ and the doped compound BaFe$_{1.8}$Co$_{0.2}$As$_2$, additional computations were performed on a unit cell of BaFe$_2$As$_2$ compressed along the $c$ axis by 0.25\%. The slight compression of the $c$-axis induced almost no change in the relaxed fractional As position, $z_{\rm As} = 0.34514$. The purpose of this calculation was to study the effect of geometrical changes of the structure on the phonons, without changing the composition.

\begin{figure}[tbp]
\center
\includegraphics[width = 7.0 cm]{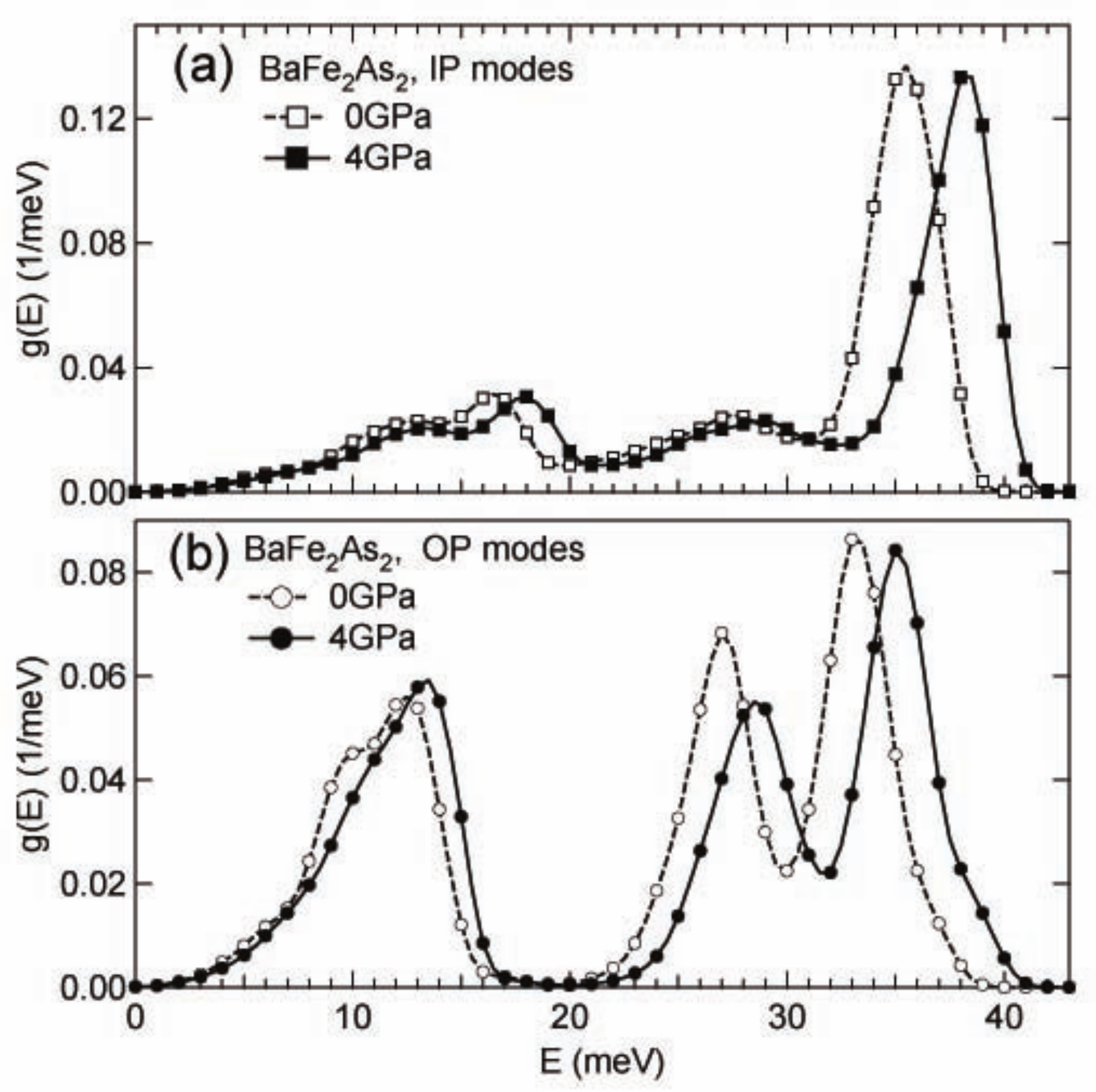} 
\caption{(a) Fe-partial DOS computed with DFT, for in-plane (IP) polarizations in BaFe$_2$As$_2$ at 0$\,$GPa and $4\,$GPa, convolved with the experimental resolution function. (b) Same for out-of-plane polarizations (OP). } \label{DFT_pressure}
\end{figure}

Results for the Fe-partial DOS for in-plane and out-of-plane polarizations at $0\,$GPa and $4\,$GPa are shown in Fig.~\ref{DFT_pressure}. The computed phonon DOS curves were convolved with the measured experimental resolution function to compare with the experimental DOS. The computed phonon DOS curves are in good agreement with the measurements, for both the shape of the DOS and the energies of the modes. The number of peaks and their relative intensities agree well with our experimental data. The DFT frequencies are consistently about 10\% stiffer than the measured frequencies, however. This shortcoming is well known for non-spin-polarized calculations \cite{Zbiri,Yildirim, Fukuda}. A treatment with spin-polarized wavefunctions typically brings better agreement with measured phonon frequencies \cite{Yildirim, Zbiri}, however the agreement within the non-spin-polarized calculations is sufficient to investigate trends as function of unit cell volume and geometry. From the average energy of the computed phonon DOS, we find the Gr\"{u}neisen parameters $\gamma^{\langle E \rangle}_{\rm IP} = 1.11$ and $\gamma^{\langle E \rangle}_{\rm OP} = 1.10$, while for the high-energy optic modes, we obtain $\gamma^{\rm optic}_{\rm IP} = 1.58$ and $\gamma^{\rm optic}_{\rm OP} = 1.15$. These theoretical values are systematically lower than our experimental results (see Table \ref{Gruneisen-exp}). This discrepancy indicates that the suppression in magnetism upon compression may result in an additional stiffening of phonons, which is not captured in the present computations.

\begin{figure}[tbp]
\center
\includegraphics[width = 7.0 cm]{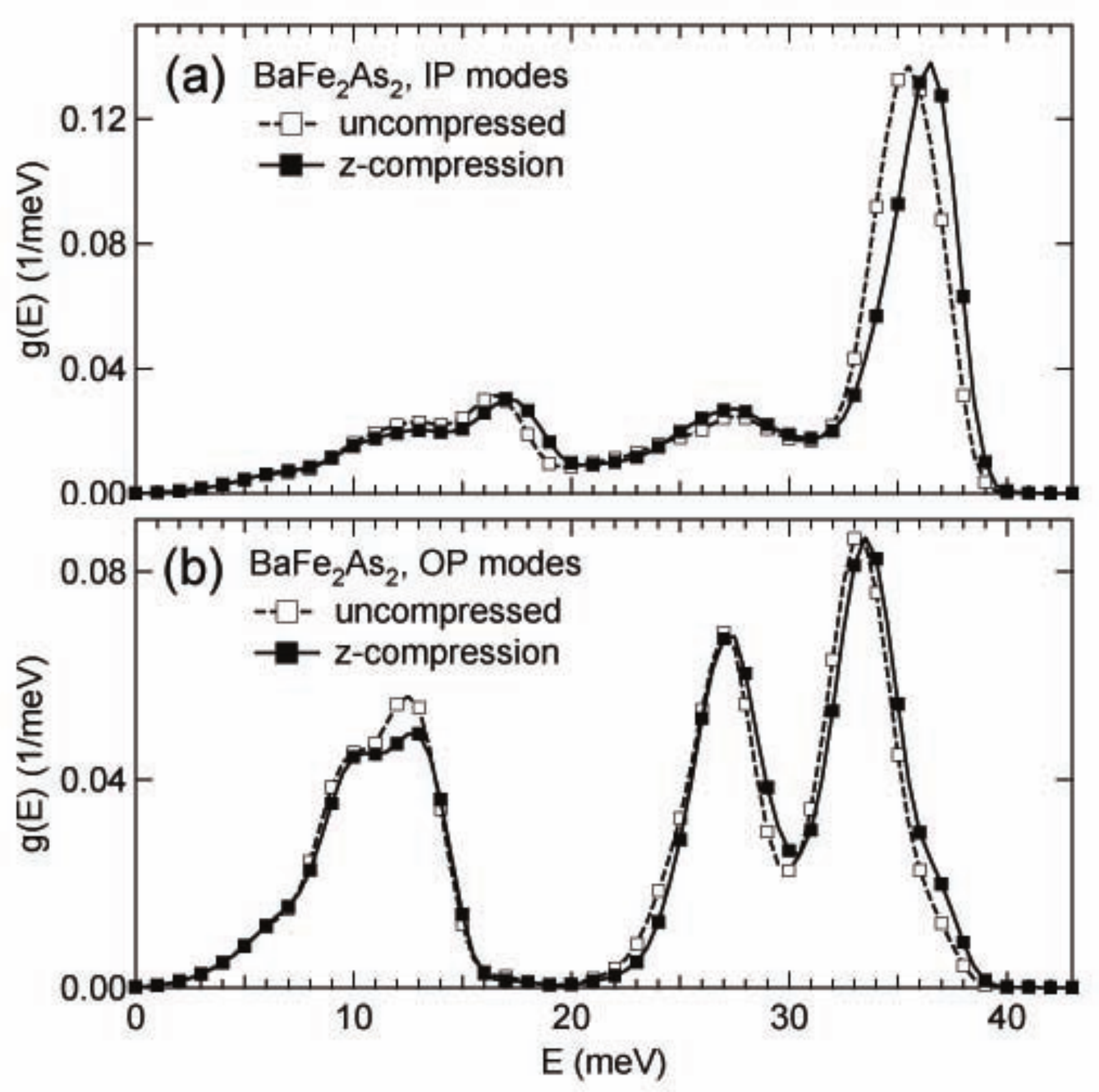} 
\caption{(a) Fe-partial DOS computed with DFT, for in-plane (IP) polarizations in BaFe$_2$As$_2$ at $300\,$K with experimental unit cell, and for BaFe$_2$As$_2$ with cell compressed $0.25\,$\% along $c$ axis. Curves are convolved with the experimental resolution function. (b) Same for out-of-plane (OP) polarizations.  } 
\label{DFT_z_compress}
\end{figure}

The projected phonon DOS computed on BaFe$_2$As$_2$ compressed along the $c$ axis is compared to the computed DOS at the experimental (uncompressed) geometry in Fig.~\ref{DFT_z_compress}. As can be seen in this figure, the compression of the $c$-axis alone leads to a stiffening of phonon modes polarized both along $c$ (OP) and perpendicular to it (IP), as observed experimentally upon Co-doping. The magnitude of the stiffening for both polarizations is also in good agreement with the experimental observation.

\section{Discussion}

Our phonon measurements as function of doping, pressure and temperature show a systematic trend. In all cases, we observe a strong correlation between the phonon energies and the unit cell volume, with contractions in the lattice parameters leading to stiffer phonons. This behavior is expected on the basis of inter-atomic force-constants weakening with increasing atomic separation. The change in phonon frequencies is related to the change in volume through the Gr\"uneisen parameter (quasiharmonic approximation), and we have determined the Gr\"uneisen parameters of IP and OP polarizations, for both the average phonon energy and the higher-energy optical peak. The Gr\"uneisen parameters obtained from measurements under pressure (listed in Table~\ref{Gruneisen-exp}) represent the dependence of phonon energy levels on pressure alone (at $300\,$K). On the other hand, the Gr\"uneisen parameters derived from our $T$-dependent measurements and the reported thermal expansion combine the effect of the volume change and other $T$-induced phenomena, such as carrier excitation across the Fermi level, or $T$-dependent magnetic excitations. 

	Upon closer examination, the contraction of the unit cell with Co doping (0.25\% in volume) at $300\,$K is expected to result in a stiffening of the average phonon energy of about 0.4\%, $\Delta \langle E \rangle / \langle E \rangle \sim 0.004$, based on our measured Gr\"uneisen parameters. The observed stiffening is in good agreement with this result, with $\Delta \langle E_{\rm OP} \rangle / \langle E_{\rm OP} \rangle = 0.004\pm0.002$ (beamline 16-IDD) and $0.006\pm0.002$ (beamline 3-IDD), and $\Delta \langle E_{\rm IP} \rangle / \langle E_{\rm IP} \rangle = 0.01\pm0.004$ (beamline 16-IDD). Thus, we conclude that electron doping results in a stiffening of phonons that can be accounted for based on the volume contraction alone. This is in agreement with the results of the DFT simulations, which show that a 0.25\% compression in the $c$ lattice parameter accounts for the observed change of both the OP and IP-polarized phonon DOS upon doping (see Fig.~\ref{DFT_z_compress}). In both theory and experiment, the peak at $32\,$meV appears to be the most sensitive to compressions of the $c$ axis (for IP as well as OP polarizations).
	
	Recently, Kimber \textit{et al.} \cite{Argyriou-pressure} have reported a similarity between the effects of pressure and doping on the lattice constants of (Ba,K)Fe$_2$As$_2$. Our measurements indicate that doping and compression also have similar effects on the phonons, which is easily understood as Co exerting an effective chemical pressure (Co is a smaller species than Fe), which leads to a smaller unit cell and stiffer phonons. 

The Gr\"uneisen parameters obtained from first-principles calculations (corresponding to lattice compressions) are significantly smaller than the experimental values determined from high-pressure phonon data. Although the non-spin-polarized calculations are expected to predict stiffer phonon frequencies than the experimental values, this should be the case at all volumes, and does not account for the underestimated Gr\"uneisen parameter. A possible explanation is that compression leads to a weakening of the magnetic moments in the material, and an extra phonon stiffening, compared to a pure volume effect. Since our calculations are non-magnetic at all volumes, this extra stiffening would be missed, leading to smaller predictions for the Gr\"uneisen parameters.

Using a recently formulated Landau theory \cite{Egami-spin-phonon}, this effect can be estimated for the case of the As Raman phonon mode, which corresponds to As planes vibrating against the Fe planes.  We may write the magnetic free energy as:

\begin{eqnarray}
{F_{\rm M} } & =  & {A M^2 + B M^4 - ( \alpha (z-z_c) - \beta (z-z_c)^2 ) M^2 } \nonumber \\
{} & & { + \frac{K_0}{2}(z- \langle z \rangle )^2 } \quad ,
\label{magnetic_free_energy}
\end{eqnarray}

\noindent where $M$ is the Fe moment, $z$ is the separation between Fe and As planes, $z_c$ is the Stoner quantum critical point in $z$ for onset of magnetism, $\langle z \rangle$ is the thermal average of $z$, and $K_0$ is the (harmonic) elastic constant without the spin-lattice interaction.  By minimizing Eq.~1 
($A$=0 at $z_c$) , we obtain:  
\begin{equation}
M = \bigg\{ \frac {\alpha (z - z_c) - \beta (z-z_c)^2} {2 B} \bigg\}^{1/2} \quad  ,
\label{moment_vs_z}
\end{equation}
\noindent which agrees with calculation \cite{Egami-spin-phonon} and experiment \cite{Egami-2}. Also, by expanding Eq.~1  
in $z$ we obtain the elastic constant renormalized for the spin-lattice interaction, 
\begin{equation}
K = K_0 \bigg\{  1 - \frac{\alpha^2}{2 K_0 B} \big[  1 + \frac{6 \beta}{\alpha} (z - z_c) \big]  \bigg\}  \quad .
\label{magnetic_free_energy}
\end{equation}
Using the values evaluated by the DFT calculation for Ba(Fe$_{0.92}$Co$_{0.08}$)$_2$As$_2$ \cite{Egami-spin-phonon}, 
$z_c = 1.278 \, {\rm \AA}$, 
$ \alpha / 2B = 11.67\, \mu_{\rm B}^2 / {\rm \AA}$, 
$\alpha = 0.193 \, {\rm eV/  \AA} \mu_{\rm B}^2$, 
$\beta = 0.137 \, { \rm eV/ \AA}^2 \mu_{\rm B}^2$
, and $\langle z \rangle  = 1.36 \, \rm \AA$,  we obtain the enhancement to the Gr\"uneisen constant due to the loss of magnetism under increasing pressure, 
\begin{equation}
\Delta \gamma = -\frac{1}{2} \frac{d \ln K}{d \ln z } = 1.81 \, .
\end{equation}
The difference in the Gr\"uneisen parameter  obtained from DFT calculations without the spin-lattice effect and the values measured by NRIXS for BaFe$_{1.84}$Co$_{0.16}$As$_2$ is $\Delta \gamma \sim 0.8$, about a factor of two smaller than the value derived above. However, this overestimate is expected since the measured value of $\gamma$ is an average over many phonon modes, while the model calculation was for the phonon mode which is most strongly affected by spin-phonon coupling. Also, the Fe-As bond is hard, and compresses less than the lattice parameters under pressure \cite{Argyriou-pressure}.

	The temperature dependence of phonon energies is compatible with the increase in volume upon thermal expansion, as was shown in Fig.~\ref{Eavg_vs_T}. However, the isobaric Gr\"uneisen parameter for high-energy optic modes (obtained from our data and reported thermal expansion curves), $\gamma|_P \simeq 2.6$,  ($P=0\,$GPa) is significantly larger than its isothermal ($T=300\,$K) counterpart, $\gamma|_T \simeq 1.6$. It thus appears that $T$-induced modifications of the electronic and/or magnetic structure may contribute directly to the $T$ dependence of phonons. It is possible that a weakening of magnetism with increasing temperature causes an enhancement in $\gamma$, similarly to the case of high pressure discussed above.
	
	Also, since the electronic DOS of BaFe$_2$As$_2$ has a  large peak in the DOS below the Fermi level, it is possible that increased temperature leads to extra free carriers, yielding larger screening and softer phonons than would be obtained through a pure volume expansion.  We point out that the $T$-dependence of phonons can exhibit systematic departures from the quasiharmonic model, when the Fermi level is in proximity of sharp features in the electronic DOS \cite{Delaire-PRL, Delaire-PRB}. In particular, metals whose Fermi level lies in a pseudo-gap in proximity of a peak in the electronic DOS, as is the case in the FeAs compounds \cite{Athena-PRL} (and also in Cr and Mo) can exhibit an excess phonon softening with increasing temperature. The dependence of phonons on temperature in Ba(Fe$_{1-x}$Co$_x$)$_2$As$_2$ thus presents some similarity with that of Cr and Mo, although this dependence is not altered over the small Co doping range which induces the superconductivity in Ba(Fe$_{1-x}$Co$_x$)$_2$As$_2$.

\section{Conclusions}

Using NRIXS, we have measured the partial phonon DOS for in-plane and out-of-plane vibrations of $^{57}$Fe atoms in BaFe$_2$As$_2$ and BaFe$_{1.8}$Co$_{0.2}$As$_2$, as function of temperature and pressure. These experimental phonon DOS curves were compared to density functional calculations. Good agreement was observed in the trend as function of doping and pressure between the experimental and theoretical results. The dependence of phonons on doping, pressure, and temperature can be understood primarily based on the dependence of the lattice constants on these thermodynamic  parameters. However, magnetism appears to induce an additional phonon stiffening upon increasing  pressure. Also, the dependence of phonons on temperature is more pronounced than would be expected for a pure volume thermal expansion, for both the parent and doped compounds. This extra softening may be attributed to thermally-induced rearrangements of electronic or magnetic structures.

\section{Acknowledgements}

We thank Steve Nagler for help with providing the isotopically-enriched iron. Experimental work at ORNL was supported by the Scientific User Facilities Division and the Division of Materials Sciences and Engineering, Office of Basic Energy Sciences, DOE. Theory work at ORNL was supported by DOE, Division of Materials Sciences and Engineering. Part of the research was performed by a Clifford G. Shull Fellow (O.D.) and by Eugene P. Wigner Fellows (A.S.S. and M.A.M.) at ORNL.
Use of the HPCAT facility was supported by DOE-BES, DOE-NNSA (CDAC), NSF, DOD TACOM, and the W.M. Keck Foundation. Use of the APS was supported by DOE-BES, under Contract No. DE-AC02-06CH11357.

\end{document}